\documentclass[conference]{IEEEtran}
\IEEEoverridecommandlockouts

\usepackage{cite}
\usepackage{amsmath,amssymb,amsfonts}
\usepackage{algorithmic}
\usepackage{graphicx}
\usepackage{textcomp}
\usepackage{xcolor}
\usepackage[caption=false,font=footnotesize]{subfig}
\def\BibTeX{{\rm B\kern-.05em{\sc i\kern-.025em b}\kern-.08em
    T\kern-.1667em\lower.7ex\hbox{E}\kern-.125emX}}
\begin{document}

\title{Integrated Optical Receiver for Communication and Fine Tracking in Inter-Satellite Links}

\author{
\IEEEauthorblockN{
Meysam~Ghanbari\IEEEauthorrefmark{1},
Mohammad~Taghi~Dabiri\IEEEauthorrefmark{2},
Akram~Alomainy\IEEEauthorrefmark{3},
Qammer~H.~Abbasi\IEEEauthorrefmark{4},\\
Mazen~O.~Hasna\IEEEauthorrefmark{5},
and~Khalid~A.~Qaraqe\IEEEauthorrefmark{1},
}

\IEEEauthorblockA{
\IEEEauthorrefmark{1}College of Science and Engineering, Hamad Bin Khalifa University, Doha, Qatar.\\
}

\IEEEauthorblockA{
\IEEEauthorrefmark{2}Department of Engineering, University of Cambridge, Cambridge, UK.\\
}

\IEEEauthorblockA{
\IEEEauthorrefmark{3}School of Electronic Engineering and Computer Science, Queen Mary University of London, London, UK.\\
}

\IEEEauthorblockA{
\IEEEauthorrefmark{4}James Watt School of Engineering, University of Glasgow, Glasgow, UK.\\
}

\IEEEauthorblockA{
\IEEEauthorrefmark{5}Department of Electrical Engineering, Qatar University, Doha, Qatar.\\
Email: megh89467@hbku.edu.qa
}
}

\maketitle

\begin{abstract}
Inter-satellite optical links demand high-precision fine tracking while preserving sufficient received power for data communication, yet these functions are often treated separately at the receiver. This paper proposes a dual-function optical receiver that integrates data reception and fine tracking on a shared, intentionally defocused receiver plane. The architecture combines a central data lens with an annular four-segment tracking detector, creating a fundamental tradeoff between data-power collection and angular-estimation capability. A scalar Fresnel wave-optical model is developed together with nonlinear two-dimensional calibration and a noise-aware worst-case angular-accuracy framework. The receiver geometry is then jointly optimized to maximize the guaranteed fine-tracking range subject to a minimum data-power constraint. Results demonstrate that appropriate co-design of the central aperture and defocus substantially enlarges the usable fine-tracking region while maintaining the required communication-path power. The proposed framework provides a receiver-level benchmark for analyzing and designing integrated communication-and-tracking architectures in future optical inter-satellite terminals.
\end{abstract}


\section{Introduction}

Inter-satellite optical links (OISLs) offer large bandwidth, narrow beam divergence, and immunity to radio-frequency interference, making them attractive for high-capacity space networks. However, narrow beams also make them highly sensitive to pointing, acquisition, and tracking (PAT) errors \cite{Dabiri2025OAMFSO}. Practical PAT systems therefore use coarse and fine stages: coarse tracking reduces initial pointing uncertainty, while fine tracking estimates the residual angle of arrival (AoA) more precisely. This residual AoA must remain within the angular region in which the fine tracker meets the required estimation accuracy \cite{Shang2025OISL,Ghanbari2026NarrowBeams}.
Fine tracking commonly uses quadrant or segmented photodetectors and differential optical-power measurements to estimate angular displacement. Communication instead favors high-sensitivity detectors such as avalanche photodiodes (APDs), whereas PIN-based segmented detectors are attractive for tracking because they avoid avalanche-gain variations and channel-matching sensitivity. Some terminals therefore separate data and tracking into two optical branches using a beam splitter or related routing element. This increases component count, alignment and packaging complexity, and divides received power between the functions, motivating an integrated receiver architecture \cite{Safi2021BeamTracking,Chen2025PointingTracking}.

Recent studies have approached the coupling of optical communication and tracking from several directions. 
In \cite{Kim2023VISION}, a nanosatellite laser-crosslink terminal shares a common front-end aperture but divides the received power between separate APD and quadrant-cell branches, while \cite{Park2022TrackingEfficiency} experimentally demonstrated a common-path FSO terminal that still uses a 7:3 beam splitter for APD data reception and QPD tracking.
To avoid this penalty, \cite{Mai2023BeaconlessAoA} proposed an integrated focal-plane receiver comprising a central data photodetector surrounded by beam-position sensing elements, eliminating the conventional tracking/data beam splitter and improving both received data power and AoA tolerance. From the tracking perspective, \cite{Takamoto2025DefocusAware} showed that intentional QD defocus can enlarge the linear tracking field of view, while introducing a tradeoff between angular range and estimation accuracy due to the resulting nonlinear detector response. More recently, \cite{Safi2025CubeSatEnabled} investigated joint data detection and fine beam tracking for CubeSat FSO links using an APD array, enabling the same detector array to support communication and beam-position estimation.
Complementary multi-aperture FSO receivers have
also exploited spatially resolved quad-detector measurements
for joint estimation of AoA, transmitter pointing error, and
channel impairments \cite{Dabiri2026HierarchicalDL}.

These studies demonstrate the benefits of shared optical paths, integrated sensing and data reception, detector-array tracking, and intentional defocus; however, receiver-design variables are generally optimized for either communication or tracking rather than as a coupled optical problem. In an integrated receiver with a central data path and surrounding segmented fine-tracking detector, enlarging the central aperture improves data collection but reduces the active tracking region, while defocus redistributes irradiance and changes tracking sensitivity and usable angular range. A receiver-level framework is therefore needed to capture this coupling through wave-optical propagation, nonlinear two-dimensional calibration, noise-limited angular accuracy, and joint geometry optimization under a data-power constraint.

Accordingly, this paper proposes an integrated dual-function OISL receiver in which a central data lens and an annular four-segment tracking detector share an intentionally defocused receiver plane. The framework combines scalar Fresnel propagation, nonlinear two-dimensional calibration, noise-aware worst-case angular RMSE, and joint geometry optimization under a minimum data-power constraint. The main contributions are threefold:
i) a compact integrated optical architecture that supports data reception and fine tracking without conventional receiver-side data/tracking beam-splitter branches; ii) a two-dimensional wave-optical tracking framework that accounts for the full segmented geometry, intentional defocus, nonlinear calibration, and detector noise; and iii) a tracking-centric receiver optimization that directly quantifies the maximum residual AoA tolerated from the coarse stage while maintaining the required fine-tracking accuracy and data-power constraint.

\section{System Model and Receiver Architecture}

The proposed receiver is modeled from the output of the coarse-acquisition stage onward; transmitter-to-receiver propagation and coarse-pointing dynamics are therefore excluded. Its input is characterized by the optical power $P_r$ incident on the primary aperture and the residual AoA components $\theta_x$ and $\theta_y$, while $D_c$ and $\Delta z$ are the receiver design variables.
Fig.~1 summarizes the proposed integrated receiver architecture and the associated coordinate systems. The residual AoA components \(\theta_x\) and \(\theta_y\) introduce a two-dimensional wavefront tilt at the primary receiver lens of diameter \(D_R\) and focal length \(f\). The integrated receiver is positioned at $
z_R = f - \Delta z$,
where \(\Delta z > 0\) denotes displacement toward the primary lens relative to the nominal focal plane. The receiver consists of a central circular data lens of diameter \(D_c\) surrounded by four annular tracking segments \(Q_1\)--\(Q_4\). A radial guard gap of width \(g_c\) separates the central lens from the active tracking annulus, while a cross-shaped inactive gap of width \(g\) separates the tracking segments. Accordingly, \(D_c\) determines the central collection aperture and the inner extent of the tracking region, whereas \(\Delta z\) controls the spatial optical distribution incident on the integrated receiver.
Let \((\xi,\eta)\) denote the transverse coordinates in the primary-lens plane and \((x,y)\) those in the integrated receiver plane. For a circular primary lens of diameter \(D_R\), the aperture radius is $
R_R = \frac{D_R}{2}$,
and the corresponding aperture function is
\begin{equation}
A_R(\xi,\eta)
=
\begin{cases}
1, & \xi^2+\eta^2 \leq R_R^2,\\
0, & \text{otherwise}.
\end{cases}
\label{eq:primary_aperture_function}
\end{equation}

\begin{figure}[t]
    \centering
    \includegraphics[width=\columnwidth]{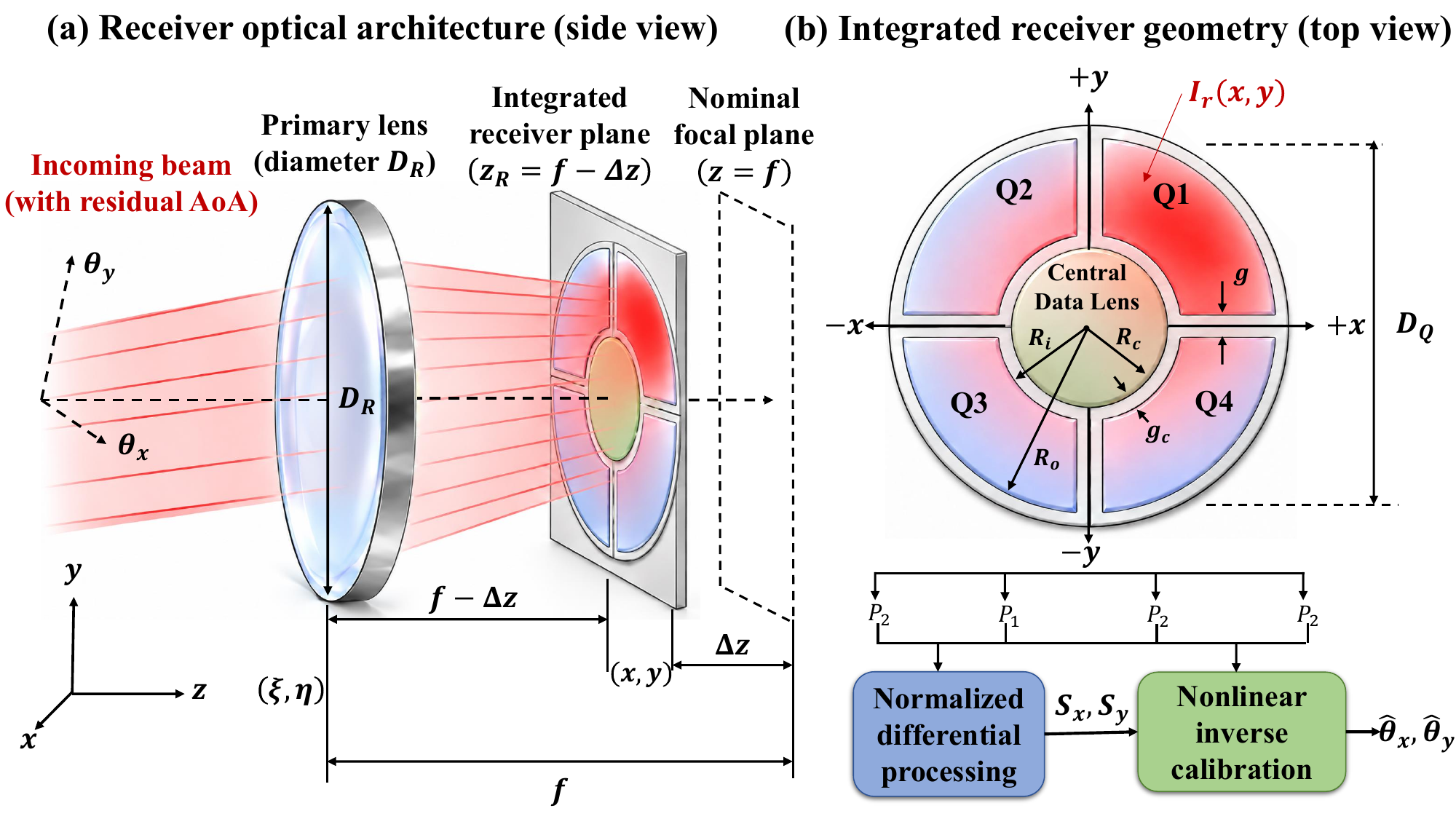}
\caption{Proposed dual-function optical receiver: (a) side view showing residual-AoA incidence and the intentionally defocused receiver plane at $z_R=f-\Delta z$; (b) top-view geometry with a central data lens, annular four-segment tracker $Q_1$--$Q_4$, and nonlinear AoA-estimation chain.}
    \label{fig:fig1}
\end{figure}

Following coarse acquisition, the field incident on the primary aperture is modeled as a locally planar tilted wavefront. For optical wavelength \(\lambda\) and wavenumber \(k=2\pi/\lambda\),
\begin{equation}
U_{\mathrm{in}}(\xi,\eta)
=
A_0
\exp\!\left[
jk\left(\theta_x\xi+\theta_y\eta\right)
\right].
\label{eq:incident_field}
\end{equation}
where \(\lvert U\rvert^2\) represents optical irradiance in \(\mathrm{W/m^2}\). For uniform pupil illumination, the field amplitude is
$
A_0
=
\sqrt{\frac{P_r}{\pi R_R^2}},
$
such that
$
\iint_{\mathbb{R}^2}
A_R(\xi,\eta)
\left|U_{\mathrm{in}}(\xi,\eta)\right|^2
\,d\xi\,d\eta
=
P_r$.
The primary receiver is modeled as an ideal thin lens of focal length \(f\) and optical power transmission coefficient \(\eta_R\). The complex field immediately after the lens is
\begin{equation}
U_L(\xi,\eta)
=
\sqrt{\eta_R}\,
U_{\mathrm{in}}(\xi,\eta)
A_R(\xi,\eta)
\exp\!\left[
-j\frac{k}{2f}
\left(\xi^2+\eta^2\right)
\right].
\label{eq:post_lens_field}
\end{equation}

The integrated receiver plane is located at \(z_R=f-\Delta z\). Under scalar paraxial Fresnel propagation, the complex field at this plane is
\begin{equation}
\begin{aligned}
U_T(x,y;z_R)
&=
\frac{e^{jkz_R}}{j\lambda z_R}
\iint_{\mathbb{R}^2}
U_L(\xi,\eta) \\
&\quad \times
\exp\!\left[
j\frac{k}{2z_R}
\left(
(x-\xi)^2+(y-\eta)^2
\right)
\right]
\,d\xi\,d\eta .
\end{aligned}
\label{eq:receiver_plane_field}
\end{equation}
The corresponding receiver-plane irradiance is
$
I_T(x,y)
=
\left|U_T(x,y;z_R)\right|^2$.
Let \(D_Q\) denote the outer diameter of the annular tracking structure. The central data-lens and outer tracking radii are
$
R_c=\frac{D_c}{2}$,
$R_o=\frac{D_Q}{2}$,
respectively, while the inner radius of the active tracking annulus is
$
R_i=R_c+g_c.
$
The central data aperture is $
\mathcal{A}_c
=
\left\{
(x,y):x^2+y^2\leq R_c^2
\right\}$.
Using the quadrant convention
$
Q_1:(+x,+y)$,
$Q_2:(-x,+y)$,
$Q_3:(-x,-y)$,
$Q_4:(+x,-y)$,
define
$
(s_{x,i},s_{y,i})
=
(+1,+1),\,
(-1,+1),\,
(-1,-1),\,
(+1,-1)$,
$
i=1,\ldots,4$.
The four active tracking regions are then expressed compactly as

\begin{equation}
\mathcal{Q}_i
=
\left\{
(x,y):
R_i^2 \leq x^2+y^2 \leq R_o^2,\;
s_{x,i}x \geq \frac{g}{2},\;
s_{y,i}y \geq \frac{g}{2}
\right\},
\label{eq:tracking_regions}
\end{equation}

for $i=1,\ldots,4$. The optical powers collected by the central data lens and the \(i\)th tracking segment are
\begin{equation}
\begin{aligned}
P_c
&=
\iint_{\mathcal{A}_c}
I_T(x,y)\,dx\,dy,
\\
P_i
&=
\iint_{\mathcal{Q}_i}
I_T(x,y)\,dx\,dy,
\qquad
i\in\{1,2,3,4\},
\end{aligned}
\label{eq:collected_powers}
\end{equation}
respectively, and the total active tracking power is
$
P_Q=\sum_{i=1}^{4}P_i$.
For power accounting, let $P_{\mathrm{dead}}$ and $P_{\mathrm{outside}}$ denote the optical powers falling on the inactive receiver regions and outside the physical receiver, respectively. With
$P_{\mathrm{plane}}=\iint_{\mathbb{R}^2} I_T(x,y)\,dx\,dy$,
power conservation gives
$P_{\mathrm{plane}}=P_c+P_Q+P_{\mathrm{dead}}+P_{\mathrm{outside}}=\eta_R P_r$.
Increasing \(D_c\) enlarges the central data aperture and, through \(R_i=R_c+g_c\), simultaneously reduces the active tracking area. In contrast, varying \(\Delta z\) changes the propagation distance and therefore the spatial irradiance distribution across both the data and tracking regions. Consequently, the central data-path power and the four tracking-segment powers are jointly dependent on \((D_c,\Delta z)\), establishing the tracking--data trade-off considered in the subsequent receiver design.

\section{Tracking Calibration and AoA Estimation}

The four tracking-segment powers \(P_1,\ldots,P_4\) are used to construct normalized differential signals for estimating the two-dimensional residual AoA. With the quadrant convention defined in Section~II, the horizontal and vertical tracking signals are
\begin{equation}
\begin{aligned}
S_x
&=
\frac{(P_1+P_4)-(P_2+P_3)}{P_Q},
\\
S_y
&=
\frac{(P_1+P_2)-(P_3+P_4)}{P_Q},
\end{aligned}
\label{eq:normalized_tracking_signals}
\end{equation}
where \(P_Q=P_1+P_2+P_3+P_4\) is the total active tracking power. For \(P_Q>0\), the normalized responses satisfy
$
-1\leq S_x,S_y\leq 1$.
Under the adopted sign convention, positive \(\theta_x\) and \(\theta_y\) produce positive local responses in \(S_x\) and \(S_y\), respectively.
Although the differential response is approximately linear near the nominal optical axis, a global linear estimator is not assumed. The finite diffraction pattern, intentional defocus, central data aperture, radial guard gap, cross-shaped inactive region, and finite tracking-detector extent result in a generally nonlinear and coupled mapping between the residual AoA and the tracking signals. 
Define the residual angular and tracking-signal vectors as
$\boldsymbol{\theta}=[\theta_x,\theta_y]^{\mathrm{T}}$
and
$\mathbf{S}=[S_x,S_y]^{\mathrm{T}}$.
The deterministic receiver response is then represented by
\begin{equation}
\mathbf{S}
=
\mathbf{F}(\boldsymbol{\theta})
=
\begin{bmatrix}
F_x(\theta_x,\theta_y)\\
F_y(\theta_x,\theta_y)
\end{bmatrix},
\label{eq:nonlinear_receiver_response}
\end{equation}
where \(F_x(\cdot)\) and \(F_y(\cdot)\) are obtained numerically from the wave-optical model in Section~II. The full two-dimensional dependence is retained; therefore, no separability between the horizontal and vertical angular responses is assumed.
The forward calibration map is generated over the numerical characterization domain
$
\mathcal{R}_{\mathrm{test}}
=
\left\{
(\theta_x,\theta_y):
\theta_{x,\min}\leq\theta_x\leq\theta_{x,\max},
\;
\theta_{y,\min}\leq\theta_y\leq\theta_{y,\max}
\right\}$.
The boundaries of \(\mathcal{R}_{\mathrm{test}}\) specify the numerical AoA sweep and do not define the usable fine-tracking range. At each angular coordinate, the receiver-plane field and irradiance are evaluated, the powers \(P_1,\ldots,P_4\) are integrated over their corresponding detector regions, and the resulting \((S_x,S_y)\) pair is stored. The resulting samples form a two-dimensional numerical calibration map.
Within a region where the forward mapping is unique and invertible, the noiseless AoA estimate is obtained as
$\hat{\boldsymbol{\theta}}=\mathbf{F}^{-1}(\mathbf{S})$,
where
$\hat{\boldsymbol{\theta}}=[\hat{\theta}_x,\hat{\theta}_y]^{\mathrm{T}}$.
The inverse mapping is constructed numerically from the sampled two-dimensional calibration data using interpolation rather than assuming a closed-form inverse, global linear gain, or prescribed polynomial response.
The local angular sensitivity of the receiver is characterized by the Jacobian matrix
\begin{equation}
\mathbf{J}(\theta_x,\theta_y)
=
\begin{bmatrix}
\dfrac{\partial S_x}{\partial \theta_x}
&
\dfrac{\partial S_x}{\partial \theta_y}
\\[6pt]
\dfrac{\partial S_y}{\partial \theta_x}
&
\dfrac{\partial S_y}{\partial \theta_y}
\end{bmatrix}.
\label{eq:tracking_jacobian}
\end{equation}

A necessary condition for local invertibility is
$\det[\mathbf{J}(\theta_x,\theta_y)]\neq 0$,
although local nonsingularity alone does not guarantee a unique global inverse. Accordingly,
$\mathcal{R}_{\mathrm{cal}}\subseteq\mathcal{R}_{\mathrm{test}}$
is restricted to coordinates for which the tracking power remains physically meaningful and
$\mathbf{F}$
remains unique, locally invertible, and nonfolding; no fixed threshold on
$P_Q$,
$\det(\mathbf{J})$,
or another sensitivity metric is imposed. Unlike a purely geometric field of view,
$\mathcal{R}_{\mathrm{cal}}$
therefore also enforces response invertibility, while the usable tracking region must further satisfy the noise-limited angular-accuracy requirement of Section~IV.
For a fixed receiver geometry \((D_c,\Delta z)\), the noiseless inverse calibration is verified by applying known AoA coordinates and comparing them with the reconstructed estimates. The corresponding numerical inversion errors are
$
e_{x,\mathrm{num}}
=
\hat{\theta}_x-\theta_x$,
$
e_{y,\mathrm{num}}
=
\hat{\theta}_y-\theta_y$.
These quantities characterize numerical errors associated with calibration-grid resolution and interpolation and are therefore distinct from the stochastic angular-estimation errors introduced by photodetection and receiver-electronics noise.

\section{Angular Accuracy and Tracking Range}

The deterministic calibration of Section III maps the residual receiver-side AoA to the normalized tracking signals. Because fine tracking follows coarse acquisition, performance is characterized by the maximum residual AoA satisfying the prescribed estimation accuracy; no statistical distribution is imposed on the residual coarse-stage pointing error.
For tracking segment \(i\in\{1,2,3,4\}\), the mean signal photocurrent is
$
\bar{i}_i=R_{\mathrm{PD}}P_i$,
where \(R_{\mathrm{PD}}\) denotes the photodetector responsivity. Let \(I_{\mathrm{bg},i}\) and \(I_{d,i}\) denote the background-induced and dark currents, respectively, \(B\) the equivalent receiver noise bandwidth, \(i_{n,i}\) the input-referred thermal-current noise density, and \(q\) the elementary charge. Under an equivalent Gaussian current-noise model, the total noise variance of the \(i\)th tracking channel is
\begin{equation}
\sigma_i^2
=
2q
\left(
R_{\mathrm{PD}}P_i
+
I_{\mathrm{bg},i}
+
I_{d,i}
\right)B
+
i_{n,i}^2 B.
\label{eq:tracking_channel_noise_variance}
\end{equation}

The deterministic dc offsets associated with background and dark currents are assumed to be characterized and removed before normalized differential processing, while their shot-noise contributions remain included in \eqref{eq:tracking_channel_noise_variance}. The resulting offset-corrected current in the \(i\)th tracking channel is modeled as
\begin{equation}
\tilde{i}_i
=
R_{\mathrm{PD}}P_i+n_i,
\qquad
n_i\sim\mathcal{N}(0,\sigma_i^2).
\label{eq:noisy_tracking_current}
\end{equation}

The noisy horizontal and vertical differential signals are formed directly from the four tracking currents as

\begin{equation}
\begin{aligned}
\tilde{S}_x
&=
\frac{
(\tilde{i}_1+\tilde{i}_4)-(\tilde{i}_2+\tilde{i}_3)
}{
\tilde{i}_1+\tilde{i}_2+\tilde{i}_3+\tilde{i}_4
},
\\
\tilde{S}_y
&=
\frac{
(\tilde{i}_1+\tilde{i}_2)-(\tilde{i}_3+\tilde{i}_4)
}{
\tilde{i}_1+\tilde{i}_2+\tilde{i}_3+\tilde{i}_4
}.
\end{aligned}
\label{eq:noisy_tracking_signals}
\end{equation}

Defining the noisy tracking-signal vector as
$\tilde{\mathbf{S}}=[\tilde{S}_x,\tilde{S}_y]^{\mathrm{T}}$,
each noisy realization is passed through the nonlinear two-dimensional inverse calibration of Section~III. For the \(m\)th realization,
\begin{equation}
\hat{\boldsymbol{\theta}}^{(m)}
=
\mathbf{F}^{-1}
\left(
\tilde{\mathbf{S}}^{(m)}
\right),
\label{eq:noisy_inverse_estimation}
\end{equation}
where the estimated AoA vector is
$\hat{\boldsymbol{\theta}}^{(m)}
=
[\hat{\theta}_x^{(m)},\hat{\theta}_y^{(m)}]^{\mathrm{T}}$.
Thus, the nonlinear calibration is applied directly to each noisy signal realization without locally linearizing the tracking response.
The corresponding componentwise angular-estimation errors are
\begin{equation}
e_x^{(m)}
=
\hat{\theta}_x^{(m)}-\theta_x,
\qquad
e_y^{(m)}
=
\hat{\theta}_y^{(m)}-\theta_y.
\label{eq:angular_estimation_errors}
\end{equation}

Because the receiver estimates a two-dimensional residual angular vector, the principal accuracy metric is the radial angular RMSE,
\begin{equation}
\mathrm{RMSE}_{\theta}
=
\sqrt{
\frac{1}{N_{\mathrm{MC}}}
\sum_{m=1}^{N_{\mathrm{MC}}}
\left[
\left(e_x^{(m)}\right)^2
+
\left(e_y^{(m)}\right)^2
\right]
}.
\label{eq:radial_angular_rmse}
\end{equation}

To characterize estimation accuracy as a function of residual pointing magnitude, the AoA is parameterized as
$
\theta_x=r\cos\phi$,
$\theta_y=r\sin\phi$,
where
$
r=\sqrt{\theta_x^2+\theta_y^2}
$
is the residual angular magnitude and \(\phi\) is its azimuth. Because the four-segment response becomes direction dependent away from the nominal optical axis, a single angular cut is insufficient to characterize the fine-tracking range. The worst-case radial RMSE at residual magnitude \(r\) is therefore defined as
\begin{equation}
\mathrm{RMSE}_{\mathrm{wc}}
\left(
r;P_r,D_c,\Delta z
\right)
=
\max_{0\leq\phi<2\pi}
\mathrm{RMSE}_{\theta}
\left(
r,\phi;P_r,D_c,\Delta z
\right).
\label{eq:worst_case_radial_rmse}
\end{equation}

The maximization in \eqref{eq:worst_case_radial_rmse} is evaluated numerically over a sufficiently dense azimuthal grid, with convergence checked with respect to the azimuthal resolution. Consequently, unfavorable orientations relative to the segment boundaries and the cross-shaped inactive region are included in the accuracy evaluation.
The required radial angular-estimation accuracy is
$\epsilon_{\mathrm{req}}=10~\mu\mathrm{rad}$. Let
$\mathcal{B}_r=\{(\theta_x,\theta_y):
\theta_x^2+\theta_y^2\le r^2\}$ denote the centered angular
disk of radius $r$. The corresponding usable fine-tracking radius, denoted by $\theta_{10}(P_r;D_c,\Delta z)$, is defined as the largest radius $r\ge0$ for which $\max_{0\le\rho\le r}\mathrm{RMSE}_{\mathrm{wc}}(\rho;P_r,D_c,\Delta z)\le\epsilon_{\mathrm{req}}$ and $\mathcal{B}_r\subseteq\mathcal{R}_{\mathrm{cal}}$.
The continuous condition over \(0\leq\rho\leq r\) prevents disconnected outer regions satisfying the accuracy requirement from being included in the usable tracking range when a smaller residual angle violates the same requirement.
Thus, \(\theta_{10}\) represents a guaranteed two-dimensional fine-tracking radius governed by the least favorable azimuthal orientation. The received optical power \(P_r\) and residual AoA are operating variables, whereas \(D_c\) and \(\Delta z\) remain receiver design variables for the joint optimization in Section~V.

\begin{figure*}[!t]
    \centering
    \subfloat[\label{fig:fig2a}]{%
        \includegraphics[width=0.47\textwidth]{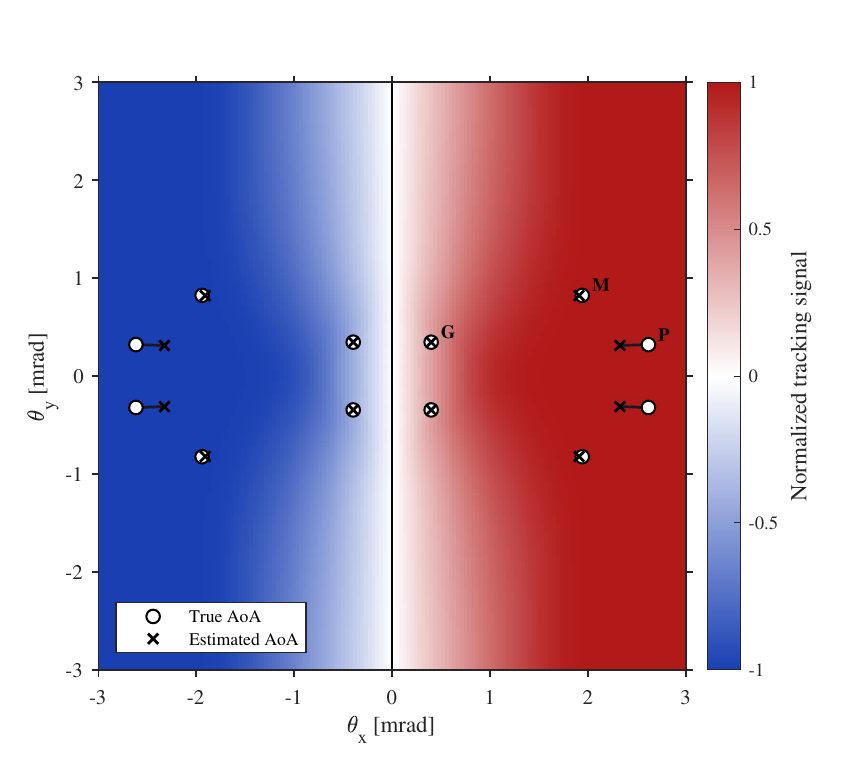}
    }
    \hfill
    \subfloat[\label{fig:fig2b}]{%
        \includegraphics[width=0.47\textwidth]{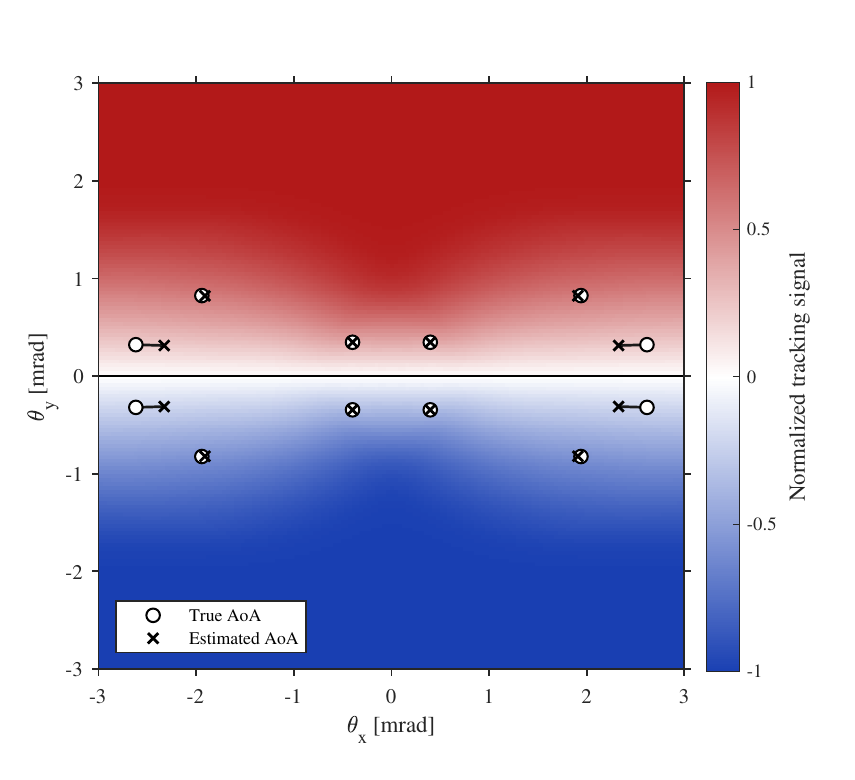}
    }
    \caption{Nonlinear 2-D calibration maps of the optimized receiver: (a) $S_x(\theta_x,\theta_y)$ and (b) $S_y(\theta_x,\theta_y)$, with true and direct-Fresnel estimated AoAs. G, M, and P denote representative good, moderate, and poor inversion cases, respectively.}
    \label{fig:fig2}
\end{figure*}

\section{Joint Tracking--Data Receiver Optimization}

The receiver geometry is jointly optimized over $D_c$ and $\Delta z$. For each candidate,
$z_R=f-\Delta z$
and
$R_i=D_c/2+g_c$,
and the complete propagation, power-integration, nonlinear-calibration, and noise-limited estimation chain is reevaluated with all remaining parameters fixed. The optimization is performed at
$P_{r,\mathrm{des}}=-40~\mathrm{dBm}=100~\mathrm{nW}$.
To maintain adequate data-path power, the central lens is required to collect at least \(25\%\) of the receiver-plane optical power at zero residual AoA:
\begin{equation}
\frac{
P_c\!\left(
0,0;
P_{r,\mathrm{des}},
D_c,
\Delta z
\right)
}{
P_{\mathrm{plane}}
\!\left(
P_{r,\mathrm{des}}
\right)
}
\geq
0.25.
\label{eq:data_power_constraint}
\end{equation}

The feasible design set is therefore
\begin{equation}
\mathcal{D}
=
\left\{
(D_c,\Delta z):
\begin{array}{l}
100~\mu\mathrm{m}
\leq
D_c
\leq
400~\mu\mathrm{m},
\\[3pt]
0.2~\mathrm{mm}
\leq
\Delta z
\leq
1.2~\mathrm{mm},
\\[3pt]
D_c/2+g_c<R_o,
\\[6pt]
\dfrac{
P_c\!\left(
0,0;
P_{r,\mathrm{des}},
D_c,
\Delta z
\right)
}{
P_{\mathrm{plane}}
\!\left(
P_{r,\mathrm{des}}
\right)
}
\geq
0.25
\end{array}
\right\}.
\label{eq:feasible_design_set}
\end{equation}

The design objective is to maximize the continuous worst-case \(10~\mu\mathrm{rad}\) fine-tracking radius defined in Section~IV. Accordingly, the optimized receiver geometry is obtained from
\begin{equation}
\left(
D_c^\star,\Delta z^\star
\right)
=
\underset{(D_c,\Delta z)\in\mathcal{D}}{\arg\max}
\;
\theta_{10}
\left(
P_{r,\mathrm{des}};
D_c,\Delta z
\right).
\label{eq:receiver_geometry_optimization}
\end{equation}

Thus, the optimization directly maximizes the guaranteed two-dimensional fine-tracking range while enforcing the minimum central data-power requirement.
The resulting optimized geometry is
$D_c^{\star}=170~\mu\mathrm{m}$,
$\Delta z^{\star}=0.450~\mathrm{mm}$,
and
$z_R^{\star}=79.550~\mathrm{mm}$
for a primary-lens focal length of
$f=80~\mathrm{mm}$.
At
$P_{r,\mathrm{des}}=-40~\mathrm{dBm}$,
the optimized geometry satisfies the data-power constraint with
$P_c/P_{\mathrm{plane}}=25.5951\%$,
yields an optical-axis
$\mathrm{RMSE}_{\theta}(0)=9.176~\mu\mathrm{rad}$,
and provides a continuous worst-case fine-tracking radius of
$\theta_{10}^{\star}=0.650~\mathrm{mrad}$.

\section{Simulation Results}
Unless otherwise stated, the simulations use an optical wavelength of \(1550~\mathrm{nm}\), a \(60~\mathrm{mm}\)-diameter primary lens with an \(80~\mathrm{mm}\) focal length, a \(1~\mathrm{mm}\)-diameter annular tracking detector, radial and cross-shaped inactive gaps of \(20~\mu\mathrm{m}\) and \(30~\mu\mathrm{m}\), respectively, and a primary-lens transmission coefficient of \(0.95\). The tracking photodetectors have a responsivity of \(0.9~\mathrm{A/W}\), with \(1~\mathrm{nA}\) dark current, \(0.5~\mathrm{nA}\) background current, \(1~\mathrm{pA}/\sqrt{\mathrm{Hz}}\) input-referred current-noise density per channel, and a \(20~\mathrm{kHz}\) noise bandwidth. A \(10~\mu\mathrm{rad}\) angular-accuracy requirement is adopted throughout the tracking-performance evaluation, while the received optical power and other parameters varied in individual studies are specified with the corresponding results.

Fig.~2 presents the nonlinear two-dimensional calibration response of the optimized receiver and verifies the inverse AoA estimator using independently evaluated direct-Fresnel test points. Fig.~2(a) shows that $S_x(\theta_x,\theta_y)$ is governed primarily by $\theta_x$, whereas Fig.~2(b) shows that $S_y(\theta_x,\theta_y)$ is governed primarily by $\theta_y$; both responses change sign about their corresponding zero-AoA axes and exhibit only weak cross-coupling near the optical axis. This agrees with the central Jacobian, whose direct sensitivities are $9.72\times10^{2}\,\mathrm{rad}^{-1}$, with negligible cross-coupling and $\mathrm{cond}(\mathbf{J}_0)=1$. Away from the optical axis, the curved transition regions reveal increasing two-dimensional coupling, while $S_x$ and $S_y$ progressively approach $\pm1$ for large $|\theta_x|$ and $|\theta_y|$, respectively, reducing local angular sensitivity. The representative G, M, and P cases quantify the resulting inversion degradation: the radial error increases from $0.048~\mu\mathrm{rad}$ at $(0.398,0.346)~\mathrm{mrad}$ to $32.0~\mu\mathrm{rad}$ at $(1.941,0.824)~\mathrm{mrad}$ and $290.9~\mu\mathrm{rad}$ at $(2.617,0.321)~\mathrm{mrad}$, where $S_x\approx0.999$. The corresponding behavior is reproduced symmetrically in all four quadrants. Moreover, the maximum direct-Fresnel/calibration-map signal mismatch is only $6.37\times10^{-5}$, confirming that the large-offset error is associated with loss of inversion sensitivity rather than numerical inconsistency.

Fig.~3 shows how the residual error left by the coarse-acquisition stage and the received optical power jointly determine the usable fine-tracking range of the optimized receiver. The horizontal dashed line at $10~\mu\mathrm{rad}$ denotes the prescribed angular-accuracy requirement, and the portion of each worst-case RMSE curve below this threshold defines the residual-AoA range over which the required accuracy is guaranteed for all azimuthal directions. At $-45~\mathrm{dBm}$, the optical-axis RMSE is already $28.56~\mu\mathrm{rad}$, so the requirement cannot be satisfied even at zero residual AoA and consequently $\theta_{10}=0$. At $-40~\mathrm{dBm}$, the axis RMSE decreases to $9.18~\mu\mathrm{rad}$, yielding a guaranteed fine-tracking radius of $\theta_{10}=0.65~\mathrm{mrad}$. Further increasing the received power expands this radius to $1.12~\mathrm{mrad}$ at $-35~\mathrm{dBm}$ and $1.33~\mathrm{mrad}$ at $-30~\mathrm{dBm}$, with corresponding axis RMSE values of $2.91$ and $0.93~\mu\mathrm{rad}$, respectively. Beyond these radii, the worst-case RMSE exceeds the requirement as the residual AoA enters the low-sensitivity regions of the nonlinear calibration response identified in Fig.~2. 
Thus, increasing received power expands the allowable coarse-stage residual AoA while the central data path remains above the prescribed $25\%$ power fraction.
The $\theta_{10}$ markers indicate the actual last-passing points of the fine radial grid rather than interpolated intersections with the $10~\mu\mathrm{rad}$ threshold.

\begin{figure}[t]
    \centering
    \includegraphics[width=\columnwidth]{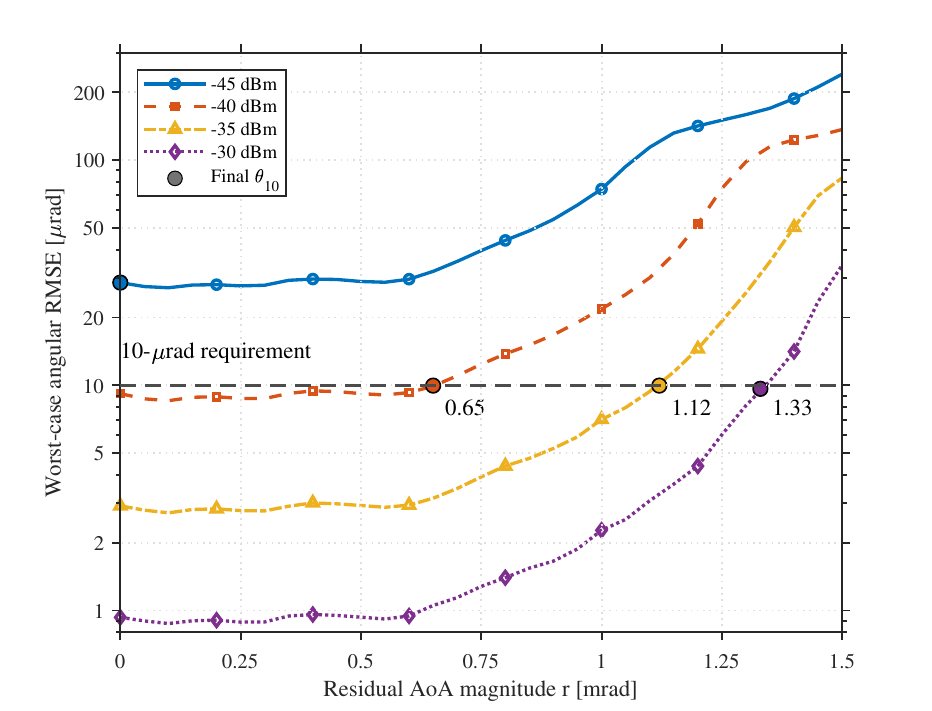}
\caption{Worst-case angular RMSE versus residual AoA magnitude at different received powers. The dashed line marks the $10~\mu\mathrm{rad}$ requirement, and filled markers denote $\theta_{10}$.}
    \label{fig:fig3}
\end{figure}

Fig.~4 summarizes the receiver-design tradeoff between central data collection and guaranteed fine-tracking range at the design power $P_{r,\mathrm{des}}=-40~\mathrm{dBm}$. The color map represents $\theta_{10}^{\mathrm{search}}$, i.e., the largest residual-AoA radius for which the worst-case angular RMSE remains below the $10~\mu\mathrm{rad}$ requirement for all smaller radii, whereas the black contour denotes the minimum acceptable central data fraction $P_c/P_{\mathrm{plane}}=25\%$. 
Within the feasible set, increasing $\Delta z$ generally improves the tracking range by redistributing more optical power toward the annular tracking detector, whereas excessive defocus reduces the central data fraction and drives the design into the infeasible region. Increasing $D_c$ produces the complementary effect: it improves central data collection but simultaneously reduces the active tracking annulus through $R_i=R_c+g_c$, thereby limiting the tracking response. The optimum therefore results from balancing these coupled mechanisms and lies close to the active data-power boundary at $(D_c^\star,\Delta z^\star)=(170~\mu\mathrm{m},0.450~\mathrm{mm})$. On the discrete search grid, this geometry yields $\theta_{10}^{\mathrm{search}}=0.610~\mathrm{mrad}$ with $P_c/P_{\mathrm{plane}}=25.5951\%$, while full-chain verification gives $\theta_{10}^{\mathrm{full}}=0.650~\mathrm{mrad}$ and an optical-axis RMSE of $9.176~\mu\mathrm{rad}$. Thus, the proposed receiver achieves its largest practical guaranteed tracking range by operating near the minimum communication-power boundary, where tracking sensitivity is enhanced without violating the required central data-power allocation.

\section{Conclusion}
This paper presented a dual-function optical receiver for OISLs that integrates data reception and fine tracking on a shared, intentionally defocused receiver plane. A wave-optical, nonlinear calibration, and noise-aware framework was developed to jointly optimize communication-power collection and tracking performance. At $-40$~dBm, the optimized design achieves a $25.60\%$ central data-power fraction, a $9.18~\mu\mathrm{rad}$ on-axis RMSE, and a guaranteed $0.65$~mrad fine-tracking radius under a $10~\mu\mathrm{rad}$ accuracy requirement. The tracking radius increases to $1.33$~mrad at $-30$~dBm. These results demonstrate the importance of joint receiver-level co-design and establish the proposed framework as a useful benchmark for integrated communication-and-tracking optical terminals.

\section*{ACKNOWLEDGMENT}
This work was supported by the Qatar Research Development and Innovation Council (QRDI) under Grant No. NPRP14C-0909-210008 and by research funding from Hamad Bin Khalifa University under the Thematic Research Grant Program Cycle 3. The statements made herein are solely the responsibility of the authors. The content is solely the responsibility of the authors and does not necessarily represent the official views of QRDI.

\begin{figure}[t]
    \centering
    \includegraphics[width=\columnwidth]{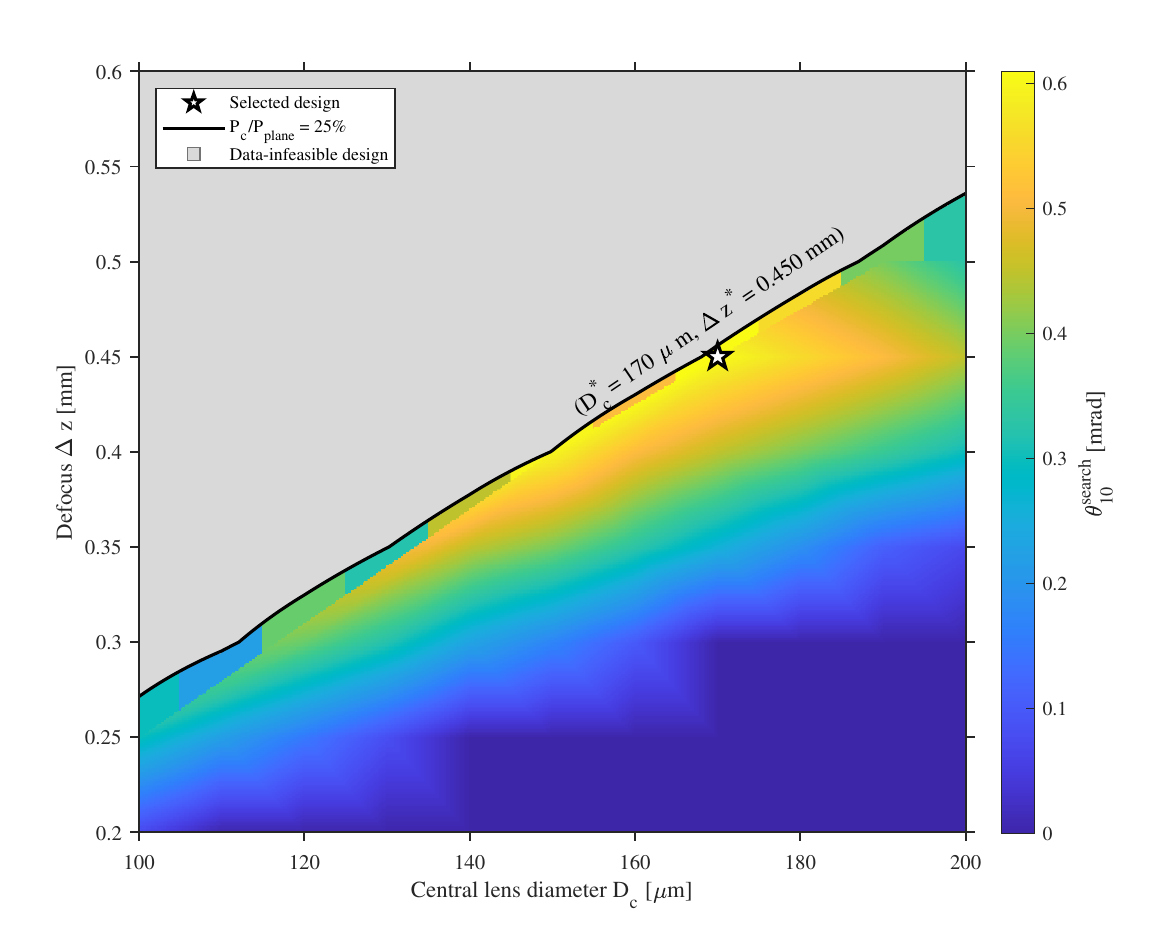}
\caption{Joint receiver optimization at
$P_{r,\mathrm{des}}=-40~\mathrm{dBm}$.
Color indicates $\theta_{10}^{\mathrm{search}}$, the black contour marks
$P_c/P_{\mathrm{plane}}=25\%$, gray denotes data-infeasible designs, and the star marks the selected design
$(D_c^{\star},\Delta z^{\star})=(170~\mu\mathrm{m},0.450~\mathrm{mm})$.}
    \label{fig:fig4}
\end{figure}

\bibliographystyle{IEEEtran}
\bibliography{myref}

\end{document}